# THE RISE AND FALL OF THE CHURCH-TURING THESIS

Mark Burgin

Department of Mathematics
*University of California, Los Angeles*
Los Angeles, CA 90095

**Abstract:** The essay consists of three parts. In the first part, it is explained how theory of algorithms and computations evaluates the contemporary situation with computers and global networks. In the second part, it is demonstrated what new perspectives this theory opens through its new direction that is called theory of super-recursive algorithms. These algorithms have much higher computing power than conventional algorithmic schemes. In the third part, we explicate how realization of what this theory suggests might influence life of people in future. It is demonstrated that now the theory is far ahead computing practice and practice has to catch up with the theory. We conclude with a comparison of different approaches to the development of information technology.

**Content**

## 1. Introduction

It looks like we are coming to the limits of Moore's law for speeding up computers – according to predictions silicon will be exhausted for further speed up by 2020. Symptomatically, experts begin to understand that, although speed is important, it does not solve all problems. As writes Terry Winograd (1997), "The biggest advances will come not from doing more and bigger and faster of what we are already doing, but from finding new metaphors, new starting points." Here, we are going to discuss one of such new starting points and to argue that it promises such possibilities for the world of computers that were unimaginable before. Moreover, it has been theoretically proved that the new approach is qualitatively more powerful than conventional computational schemes.

If we want to achieve some goal, we need to find a way to do it. Consequently, if we are looking how to increase essentially the power of computing devices, to make them intelligent and reliable, we need to find a road that will bring us to computing devices with all these properties. Here we are going to describe such a road.

Consequently, the essay has a three-fold aim. The first goal is to show how mathematics has explicated and evaluated computational possibilities sketching exact boundaries for the world of computers. It is a very complex and sophisticated world. It involves interaction of many issues: social and individual, biological and psychological, technical and organizational, economical and political. However, humankind in its development created a system of intellectual "devices" for dealing with overcomplicated systems. This system is called science and its "devices" are theories.

When people want to see what they cannot see with their bare eyes, they build and use various magnifying devices. To visualize what is situated very far from them, people use telescopes. To discern very small things, such as microbes or cells of living organisms, people use microscopes. In a similar way, theories are "magnifying devices" for mind. They may be utilized both as microscopes and telescopes. Being very complex these "theoretical devices" have to be used by experts.

Complexity of the world of modern technology is reflected in a study of Gartner Group's TechRepublic unit (Silverman, 2000). According to it, about 40% of all internal IT projects are canceled or unsuccessful, meaning that an average of 10% of a company's IT department each year produces no valuable work. An average canceled project is terminated after 14 weeks, when 52% of the work has already been done, the study shows. In addition, companies spend an average of almost $1 million of their $4.3 million annual budgets on failed projects, the study says. However, companies might be able to minimize canceled projects as well as the time for necessary cancellation if they have relevant evaluation theory and consult people who know how to apply this theory.

All developed theories have a mathematical ground core. Thus, mathematics helps science and technology in many ways. Scientists are even curious, as wrote the Nobel Prize winner Eugene P. Wigner in 1959, why mathematics being so abstract and remote from reality is unreasonably effective in the natural sciences. It looks like a miracle.

So, it is not a surprise that mathematics has its theories for computers and computations. The main of these theories is *theory of algorithms*. It explains in a logical way how computers function and how they are organized. Consequently, we are going to demonstrate how theory of algorithms evaluates computers, nets, and all computational processes suggesting means for their development.

In particular, a search for new kinds of computing resulting in elaboration of DNA and quantum computing, which are the most widely discussed. At this point, however, both these paradigms appear to be restricted to specialized domains (DNA for large combinatorial searches, quantum for cryptography) and there are no working prototypes of either. Theory of algorithms makes possible to evaluate them and find the correct place for them in a wide range of different computational schemes.

The second objective of this essay is to explain how mathematics in its development has immensely extended these boundaries opening in such a way new unexpected perspectives for the world of computers. What is impossible from the point of view of traditional mathematics and has been argued as absolutely unworkable becomes attainable in the light of the new mathematics. A new mathematical direction, which opens powerful and unexpected possibilities, is *theory of super-recursive algorithms*.

To make these possibilities real, it is necessary to attain three things:
- to develop a new approach or even a new paradigm for computing;
- to build new computers that are able to realize the new paradigm;

- to teach users (including software designers and programmers) how to utilize computing devices in a new way.

Most of what we understand about algorithms and their limitations is based on our understanding of Turing machines and other conventional models of algorithms. The famous Church-Turing Thesis claims that Turing machines give a full understanding of computer possibilities. However, in spite of this Thesis, conventional models of algorithms, such as Turing machines, do not give a relevant representation of a notion of algorithm. That is why an extension of conventional models has been developed. This extension is based on the following observation. The main stereotype for algorithms states that an algorithm has to stop when it gives a result. This is the main problem that hinders the development of computers. When we understand that computation can go on but we can get what we need, then we go beyond our prejudices and immensely extend computing power.

The new models are called super-recursive algorithms. They provide for a much more computing power. This is proved mathematically (Burgin,1988). At the same time, they give more adequate models for modern computers and Internet. These models change the essence of computation going beyond the Church-Turing Thesis (here we give a sketch of a proof for this statement) and form, consequently, a base for a new computational paradigm, or metaphor as says Winograd. Problems that are unsolvable for conventional algorithmic devices become tractable for super-recursive algorithms. The new paradigm gives a better insight into the functioning of the mind opening new perspectives for artificial intelligence. At the same time, this form of computation will eclipse the more familiar kinds and will be commercially available long before exotic technologies such as DNA and quantum computing.

The third aspire of the essay is to speculate how these new possibilities of computers might change the world, in which we live. In particular, problems of artificial intelligence are discussed in the context of human-computer interaction. Basing on new theories, it is feasible to explain what is possible to do with those computers, which we have these days. To achieve this we have to understand better the essence of modern computing. In addition to this, we discuss what do future devices that will incorporate the new paradigm for computation to a full extent.

Now we encounter a situation when a lot of forecasting of various kinds is made. However, to rely on predictions, it is necessary to distinct grounded predictions from hollow speculations.

Thus, we need to understand how people foresee and what ways of prediction are more reliable than others.

People can predict future in three ways. It is possible to use pure imagination neglecting contemporary scientific knowledge. Another extremity is to use imagination inside the cage of contemporary knowledge. As said Buddha the truth lies between two extremities. In our case, we open our way outside the cage of present-day understanding and go ahead of what we have now (for example, beyond Internet) not by pure imagination, but by the progress of scientific knowledge itself. After going beyond modern technology, we use imagination to project the consequences of the new achievements of science.

Thus, we are going along the third way basing our speculations on mathematical knowledge about computers and computations. Consequently, our futuristic picture is based not on mere dreams or up-to-date empirical inventions but on definite theoretical achievements, which, in this case, are ahead of practice. Some claim that in many spheres practice leaves theory behind and theory has only to explain what practice has already gained. It is not so with theory of algorithms. Now chemists are designing only the simplest computational units on the molecular level, while theory of parallel computations, which include molecular computing, has a lot of advanced results. Physicists are only coming to elaboration of methods for building quantum computers, while theory of quantum computing has many results, which demonstrate that it will be much more efficient that contemporary computational schemes. The same even to a greater extent is true for super-recursive algorithms. Now practice has to catch up with the theory and it is urgent to know how to bridge the existing gap.

**2. Computing through the "microscope" of theory of algorithms**

Mathematics is a powerful tool that helps people to achieve new goals as well as to understand what is impossible to do. When our experience fails, we have to go to a good theory or to elaborate such a theory. This theory, like a microscope, might allow us to see what is indistinguishable by the common sense. That is why, when computers were created, it became a task of mathematics to help to build new more powerful computers as well as to find the boundaries for power of computers. In other words, mathematics has to answer the question what computers can do and what they can't. This is the task of computer science, which includes as its nucleus theory of algorithms.

Without an adequate theory, common sense often leads people to fallacies. Even empirical knowledge, which is more advanced than common sense, may be misleading for the best experts when they ignore theory. For example, one of the best experts in artificial intelligence formulated in his book published in 1986 the "puzzle principle."

**Puzzle Principle:** *We can program a computer to solve any problem by trial and error, without knowing how to solve it in advance, provided only that we have a way to recognize when the problem is solved.*

However, when we go to theory of algorithms, we find that existence of undecidable problems has been proved there in thirties of the $20^{th}$ century. It means that any such problem cannot be solved by a computer because computer can do only what algorithms prescribe to do. Consequently, not any problem can be solved by trial and error and the "puzzle principle" fails.

It is interesting that its author formulated the "puzzle principle" in order to demonstrate that common sense can be wrong while simple empirical knowledge can remedy the situation. We see that we need more advanced knowledge, which is achievable only through theory. Without theory people repeat the same mistake many times.

This happened with the "puzzle principle." Later, in the nineties, a similar claim has been made with respect to some class of much powerful algorithms that are called trial-and error machines, which are some kind of super-recursive algorithms. They embody the puzzle principle on a higher level than conventional algorithms. However, theory shows that undecidable problems exist even for super-recursive algorithms. Consequently, the puzzle principle being very plausible, "seek and you will find", is actually invalid both in recursive and super-recursive context.

Theory of algorithms is useful not only for evaluation of validity for theoretical puzzles, but also for practical purposes. As an example, we consider the debugging problem. Everybody knows that to achieve reliable functioning of a computer, it is necessary to have correct programs. Computer crashes can result in anything from mild inconvenience to the loss of human lives, if the systems that run nuclear power stations were to malfunction, for example. However, it is almost impossible to write a program that from the beginning does not have mistakes or, as programmers call them, bugs. Thus, when written, almost all programs have bugs, and we bump into a vital debugging problem. It is a very complicated problem, and it is natural to look for a way to make a computer to debug programs by itself. In other words, it is urgent to design debugging programs. For many years, programmers tried to elaborate such programs. They suggested different theoretical and empirical methods, such as proving program correctness, or used heuristic procedures, but were not able to solve the problem of computerized debugging. Thus, we may ask a question whether theory of algorithms can help programmers in their search.

Results of this theory allow us to find an answer. To our regret, theory of algorithms states that *it is impossible to write a program that can debug any other program*. This is a consequence of the fact that the halting problem for Turing machines is undecidable.

To remedy this, theoreticians suggested using logic for proving program correctness. The reason is that logic is a powerful tool in mathematics. However, theory of algorithms enlightens us that it is impossible to prove program correctness for all programs.

Here a reader may get an impression that theory of algorithms produces only negative results. It is wrong. One of the brightest examples is its contribution to computer architecture. History of computers tells us that when the first electronic computer had been created in the USA, those who did it invited the great mathematician John von Neumann to look at it. After he was introduced to the principles of its functioning, von Neumann suggested a more advanced architecture for computer. It has been called the von Neumann architecture. For a long time all computers had the von Neumann architecture in spite that all other components of computer (elements of hardware, software, and interface) changed very rapidly. This is a well-known fact. However, very few know that the architecture suggested by von Neumann copied of the structure of one of the most popular theoretical models of algorithm, a Turing machine. Von Neumann himself did not referred to this model suggesting the new architecture, but being an expert in theory of algorithms he knew Turing machines excellently.

In our days, theory of algorithms helps to solve such vital practical problems as web reliability, communication security, computer efficiency and many others.

Now let us look a little bit deeper into theory of algorithms. Its very name indicates that algorithm is in its center. However, it is necessary to make a distinction between the informal notion of algorithm and its mathematical models. Informal notion is used in everyday life, in reasoning of experts, as well as in methodology and philosophy of computer science, mathematics, and artificial intelligence. At the same time, mathematical models constitute the core of theory of algorithms.

An informal notion of algorithms is comparatively vague, flexible, and easy to treat. Consequently, it is insufficient for an exact study. In contrast to this, mathematical models are precise, rigid, and formal. Consequently, they capture, as a rule, only some features of informal notions. Thus, to get a better representation, we need constantly to develop mathematical models. This has always been the case with all basic notions, which mathematics acquired from the real world. For example, the notion of number gave birth to a series of mathematical concepts: from natural to rational and integer to real to complex numbers to hypernumbers and transfinite numbers. The same is true for the notion of algorithm.

The word "*algorithm*" has an interesting historical origin. It derives from the Latin form of name of the famous Arab mathematician Muhammad ibn Musa al-Khowarizmi. He was born sometime before 800 A.D. and lived at least until 847. He wrote his main works *Al-jabr wa'l muqabala* (from which our modern word "*algebra*" comes) and a treatise on Hindu-Arabic numerals while working as a scholar at the House of Wisdom in Baghdad. The Arabic text of the latter book is lost but a Latin translation, *Algoritmi de numero Indorum,* which means in English *Al-Khowarizmi on the Hindu Art of Reckoning*, introduced to the European mathematics the Hindu place-value system of numerals based on 1, 2, 3, 4, 5, 6, 7, 8, 9, and 0. The first use of zero as a place holder in positional base notation was probably due to al-Khowarizmi in this work. Methods for arithmetical calculation were given in this book. These methods were the first that were called algorithms following the title of the book, which begins with the name of the author. In such a way, the name *Al-Khowarizmi* became imprinted into the very heart of mathematics. Now the notion of algorithm has become one of the central concepts of mathematics. It is a cornerstone of one of the approaches to the foundations of mathematics as well as of the whole computational mathematics. Moreover, the term "algorithm" became a general scientific and technological concept used in a variety of areas.

A popular point of view on algorithm is presented by Rogers (1987):

algorithm is a *clerical (i.e., deterministic, book-keeping)* procedure which can be applied to any of a certain class of symbolic *inputs* and which will eventually yield, for each such input, a corresponding *output*.

More generally, an algorithm is a specific kind of recipe, method, or technique for doing something. In other words, *an algorithm is a text giving unambiguous (definite) and simple to* *follow (effective) prescriptions (instructions) how from given inputs (initial conditions) derive necessary results*.

Algorithms are only rules, but people usually say and write that algorithms compute and solve problems. For example, algorithms of addition for natural numbers add numbers. If all values of a function can be, at least, theoretically, computed according to some algorithm, we say that the algorithm computes this function. In a similar way, we assume that algorithms solve some problems and cannot solve other problems.

Actually, algorithms are something you use every day, sometimes even without much conscious thought. When you want to know time, you look at a watch or clock. This simple rule is an algorithm. When we want to drive, we come to a car, sit down at the driver sit, fasten belts, and start engine. This is also an algorithm.

When we do calculations, we use algorithms. All calculations are performed according to algorithms that control and direct those calculations. All computers, ordinary and programmable calculators function according to algorithms because all computing and calculating programs are algorithms represented by means of programming languages. In many cases, people behavior is organized according to algorithms. We may speak, for example, about algorithms of buying some goods, products or food.

However, in real life when we encounter complex situations, algorithms become too rigid. Consequently, formalized functioning of complex systems (such as people) is mostly described and controlled by more general than algorithms structures. They are called procedures. Algorithms are such cases of procedures that may be performed by some "mechanical" devices. .

Now it is assumed that the most powerful "mechanical" devices for performing formalized data transformations are computers. Consequently, algorithms are restricted to those procedures that are realized by computers. At the same time, everything that computer can do is presented by algorithms in a form of computer programs.

However, historically algorithms had appeared long before the first computers were built. Thus, initially the connection between algorithms and computers was not assumed. For a definite time, it was supposed that any system of operations, which a person, who is equipped only with a pencil and paper, can complete, is an algorithm. Such an approach either extends the scope of algorithms far beyond the conventional limits or reduces a human being to a computer. So, we may ask which of these two cases is true. To solve this and other problems, it is necessary to have an exact mathematical concept of algorithm – an informal notion is insufficient. This was done by theory of algorithms. Formation of the conventional concept of algorithm, and thus, of computability, is one of the major achievements of the 20$^{th}$ century mathematics.

Being rather practical, theory of algorithms is a typical mathematical theory with a quantity of theorems and proofs. However, the main achievement of this theory has been the elaboration of an exact mathematical model of algorithm. It was done less than seventy years ago. First models appeared in mathematics in thirties of the 20$^{th}$ century in connection with its intrinsic dilemmas of finding solutions to some mathematical problems.

An important peculiarity of the exact concept of algorithm is that it exists in various forms. Different mathematicians suggested different models of algorithm: Turing machines (deterministic with one tape, with several tapes, with several heads, with n-dimensional tapes; non-deterministic, probabilistic, alternating, etc.), partial recursive functions, Post productions, Kolmogorov algorithms, finite automata, vector machines, register machines, neural networks, Minsky machines, random access machines (RAM), array machines, etc.

Some of these models (such as recursive functions or Post productions) give only rules for computing. Others (such as Turing machines or neural networks) also present a description of a computing device, which functions according to the given rules.

The most popular model was suggested by the outstanding English mathematician Alan Turing. Consequently, it is called a deterministic Turing machine. This abstract device consists of the three parts (cf. Fig 1): *the control unit*, which contains rules for functioning, has different states that are changed during computation, and is sometimes equalized with the Turing machine itself; *operating unit*, which is called the head of the Turing machine; and *memory*, which has the form of a potentially infinite tape (or several tapes) divided into cells. Some of the states of the control unit are called final states. The head can be at some cell of the tape (or observe its content from outside}. Each cell may contain some symbol from a given alphabet of the Turing machine or it may be empty. The head may write a symbol into the empty cell where it is situated or rewrite the symbol if the cell is not empty. Taking into account the content of the cell with the head and the state of the control device, the rules of this machine tell it what to do further.

The result of computation is determined as follows. If a Turing machine does not stop functioning, it is assumed that it gives no result. Turing machine stops functioning in two cases: either it cannot find the rule to continue or it comes to a final state. Some of the final states are resultless, while others indicate that the Turing machine has found a solution to the problem it solves.

To compare classes of algorithms, we introduce computing power and equivalence of such classes. Two classes of algorithms are equivalent (or more exactly, functionally equivalent) if they compute the same class of functions. A weaker class of algorithms has less computational power because it allows computing fewer functions than a stronger class of algorithms. For example, the class of all finite automata is weaker than the class of all deterministic Turing machines. It means that a deterministic Turing machine can compute everything that can compute finite automata. However, there are such functions that are computable by deterministic Turing machines while finite automata cannot compute them.

In spite of all differences, it has been proved that each of the mathematical models of algorithm is either weaker or equivalent to the class of all deterministic Turing machines with one tape (or equivalently, to the class of all partial recursive functions). Moreover, all modifications of deterministic Turing machines (deterministic Turing machines with several tapes, with several heads, with n-dimensional tapes; non-deterministic, probabilistic, alternating, and reflexive Turing machines, etc.), which do not include such non-algorithmic blocks as oracles, are equivalent to the class of all deterministic Turing machines with one tape.

These results makes possible to evaluate computing power of the new computing schemes. Now there are several perspective approaches how to increase the power of computers and networks. We may distinct chemical, physical, and mathematical directions. Two first are applied to hardware only influencing software and infware, while the mathematical approach transforms all three components of computers and networks. The first approach is very popular. It is called the molecular computing, the most popular branch of which is the DNA computing (Cho, 2000). Its main idea is to design such molecules that solve computing problems. The second direction is even more popular than the first. It is quantum computing (Deutsch, 2000). Its main idea is to perform computation on the level of atoms. The third direction is called theory of super-recursive algorithms (Burgin, 1999). It is based on a new paradigm for computation that changes computational procedure.

However, both first types of computing, molecular and quantum, can do no more than conventional Turing machines theoretically can do. For example, quantum computers are only some kinds of nondeterministic Turing machines, while a Turing machine with many tapes and heads model DNA and other molecular computers. DNA and quantum computers will be (when they will be realized) eventually only more efficient. In practical computations, they can solve more real-world problems than Turing machines. However, any modern computer can also solve more real-world problems than Turing machines because these abstract devices are very inefficient.

Here, it is worth mentioning such new computational model as reflexive Turing machines (Burgin, 1992). Informally, they are such machines, which can change their programs by themselves. Genetic algorithms give an example of such an algorithm that can change its program while functioning. In his lecture at the International Congress of Mathematicians (Edinburgh, 1958), the famous American logician Steven Kleene proposed a conjecture that a procedure that can change its program while functioning would be able to go beyond the Church-Turing Thesis. However, it was proved that such algorithms have the same computing power as deterministic Turing machines (Burgin, 1992). At the same time, it is proved that reflexive Turing machines can essentially improve efficiency. Besides, reflexive Turing machines illustrate creative processes facilitated by machines, which is very much on many people's minds. It is noteworthy that Hofstadter is surprised that a music creation machine can do so well because this violates his own understanding that machines only follow rules and that creativity cannot be described as rule-following.

All models of algorithms that are equivalent to the class of all deterministic Turing machines are called *recursive algorithms*, while classes of such algorithms are called Turing complete.

Those kinds of algorithms that are weaker than deterministic Turing machines are called *subrecursive algorithms*. Finite and stack automata, recursive and primitive recursive functions give examples of subrecursive algorithms.

For many years all attempts to find mathematical models of algorithms that were stronger (could compute more) than Turing machines were fruitless. This situation influenced the emergence of the famous *Church-Turing Thesis*, or it is better to say *Church-Turing Conjecture.* It states that *the informal notion of algorithm is equivalent to the concept of a Turing machine.* In other words,

**any problem that can be solved by an algorithm and any algorithmic computation that can be done by some algorithm can be solved and done by some Turing machine.**

Here we skip the problem of efficiency because conventional Turing, machines, being very simple, are very inefficient. Consequently, we leave problems of tractability beyond the scope of this paper considering only computability and computing power.

The Turing-Church Thesis is extensively utilized in theory of algorithms as well as in the methodological context of computer science. It has become almost an axiom. However, it has been always only a plausible conjecture like any law of physics or biology. It is impossible to prove such a conjecture completely. We can only add some supportive evidence to it or refute it. At the same time inside mathematics, we can prove or refute this conjecture if we choose an adequate context.

In addition to this, it is necessary to understand that all mathematical constructions that embody the informal notion of algorithm are only models of algorithm. Consequently, what is proved for these models has to be verified for real computers. In our case, we need to test whether recursive algorithms give an adequate representation for modern computers and networks and whether it is possible to build such computers that go beyond the recursive schema. We will see later that the answer to the first question is negative, while the second problem has a positive solution.

### 3. New perspectives through the "telescope" of theory of algorithms

Being able to explain a lot about what is going on, theory can also help us, like a telescope, to see far ahead of us. However, theory has these abilities only when its achievements go ahead of practice. Theory of super-recursive algorithms has this potency.

To understand, at least, something about this theory, we need to return to the Turing-Church Thesis and to explain that it was refuted when *super-recursive algorithms* appeared. The first super-recursive algorithms were introduced in 1965. Two American mathematicians Mark Gold and Hillary Putnam brought in concepts of limit recursive and limit partial recursive functions. Their papers were published in the same issue of the Journal of Symbolic

Logic, although Gold had written about these ideas before. It is worth mentioning that constructions of Gold and Putnam were rooted in the ideas of non-standard analysis originated by Abraham Robinson (1966) and inductive definition of sets (Spector, 1959). As a matter of fact, Gold was a student of Robinson.

Ideas of Gold and Putnam gave birth to a direction that is called inductive inference (Gasarch and Smith, 1997) and is a fruitful direction in machine learning and artificial intelligence.

Limit recursive, limit partial recursive functions and methods of inductive inference are super-recursive algorithms and as such can solve such problems that are unsolvable by Turing machines. Although, being in a descriptive and not constructive form, they were not accepted as algorithms for a long time. Even introduction of a device that were able to compute such functions (Freyvald, 1974) did not change the situation. Consequently, this was an implicit period of the development of theory of super-recursive algorithms.

In 1983 the author independently of inductive inference and limit recursion introduced inductive Turing machines that included all previous models of algorithms. From the beginning, inductive Turing machines were treated as algorithms. Thus, it was not by chance that their implications for the Church-Turing Thesis and the famous Gödel incompleteness theorem were considered (Burgin, 1987) refuting the Thesis and changing understanding of the theorem. This was the beginning of the explicit stage for theory of super-recursive algorithms.

To understand the situation, let us look at the conventional models of algorithm. We can see that an extra condition appears in formal definitions of algorithm, that is, after giving a result algorithm stops (cf., for example, Harel, 2000). It looks natural for what you have to do more after you have what you wanted. However, if we analyze attentively what is going on with real computers, we have to change our mind.

Really, no computer works without an operating system. Any operating system is a program and any computer program is an algorithm according to the general understanding. At the same time, a recursive algorithm has to stop to give a result, but we cannot say that a result of functioning of operating system is obtained when computer stops functioning. On the contrary, when computer is out of service, its operating system does not give the necessary result. Moreover, any operating system does not produce a result in a form of some word, while this is an essential condition for any recursive algorithm. Although, from time to time, operating system sends some messages (strings of words) to a user, the real result of operating system is reliable functioning of the computer. Stopping when the computer is shut down is only a partial result. Consequently, the result of the operating system functioning is obtained only when computer does not stop (at least, potentially). Other similar results are considered in (Burgin, 1999). Thus, we come to a conclusion that it is not necessary for an algorithm to stop after getting a result.

So far, so good, but how to determine a result when the algorithm does not stop functioning?

Mathematicians found an answer to this question. Moreover, a result of non-stopping computation may be defined in different ways. Here we consider the simplest case realized by inductive Turing machines.

In a structured way (Burgin, 1983), a Turing machine **M** is represented by a triad (**H, Q, K**) where **H** is the object domain of **M**, **Q** is the state domain of **M,** and **K** is the memory domain, or the structured memory of **M**. All these domains are structured. However, we are not going to describe these structures in order not to make this text too difficult for comprehension. We give only short informal description.

In the simplest case, a Turing machine **M** works with words in some alphabet *A* (domain **H**) and has an operating device **h**, which is called the head of **M** and is a part of **Q.** Memory of **M** is a tape divided into cells. In each cell, a symbol from the alphabet *A* are written. The head **h** can, move from cell to cell, read and change these symbols according to rules of **M**, which constitute a part of **Q**.

A model of a Turing machine, which is more relevant to computers, is given in the Figure 1.

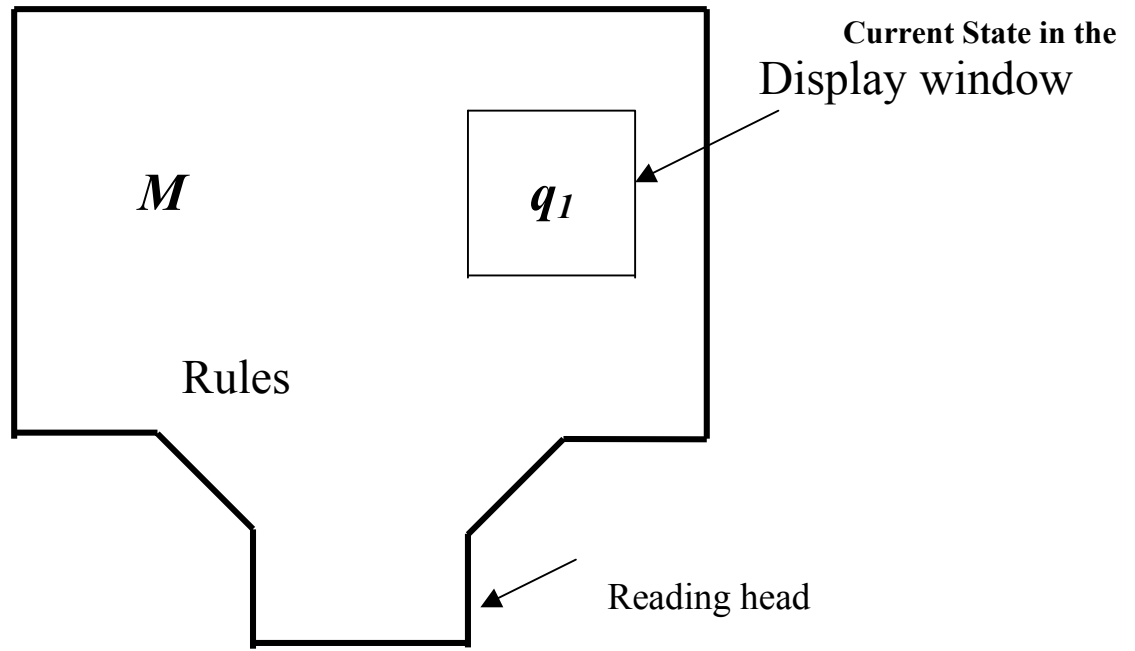

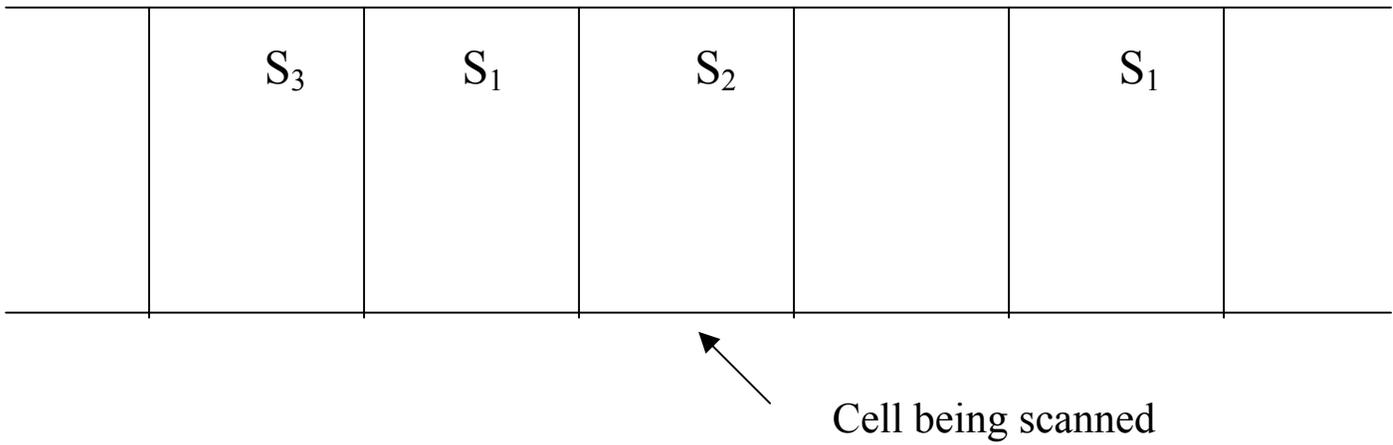

**Fig. 1** Turing machine with one moving tape and one static head

In a similar way, the simplest realistic inductive Turing machine has the same structure as a conventional Turing machine with three tapes and three heads: input, working, and output tapes and heads. Both, inductive and ordinary Turing machines make similar steps of computations. The difference is in the determination of their output. We know (cf. section 3) that a conventional Turing machine produces a result only when it halts. We assume that this result is a word written on the output tape. Such simple inductive Turing machine also produces words as its results. In some cases, an inductive Turing machine stops in a final state and gives a result like a conventional Turing machine. The difference begins when the machine does not stop. An inductive Turing machine can give a result without stopping. To show this, we consider the output tape and assume that the result has to be written there.

It is possible that in the sequence of computations after some step, the word that is written on the output tape is not changing in spite that the machine continues its work. This word, which is not changing, is taken as the result of this computation. Thus, an inductive Turing machine does not halt but produces a definite result after a finite number of computing operations. It explains the name "inductive" as in induction we go step by step checking if some statement is true for an unlimited sequence of cases.

While working without halting, an inductive Turing machine can occasionally change its output as it computers more. However, human beings are not put off by a machine that occasionally changes its outputs (as in "clock paradigm", which is considered in the next section). They can be satisfied that the result just printed is good enough, even if another (possibly better) result may come in the future. And if you continue your computation, it will eventually come. Another example is a program that outputs successively better approximations to a number a user is interested in; after a few digits of accuracy are attained, she or he can use the output generated even if the machine is not "done".

To show that inductive Turing machines are more powerful than ordinary Turing machines, we need to find a problem that no ordinary Turing machine can solve and to explain how some inductive Turing machine solves this problem. To do this, let us take the problem, which was found one the first to be unsolvable and now is one of the most popular in the theory of algorithms. This is the halting problem for an arbitrary Turing machine with a given input. It was proved by Turing that no Turing machine can solve this problem for all Turing machines. Here is a short outline of this proof.

Really, let us consider all Turing machines that work with words in the alphabet {1,0} and suppose that there is such a Turing machine **A**, which solves the halting problem. From the theory of Turing machines, it is known that there are such a Turing machine **D**, which generates descriptions of all Turing machines in the alphabet {1,0}, and a Turing machine **U**, which given a description of a Turing machine, can simulate its functioning. In addition to this, we take some natural enumeration of all words in the alphabet {1,0} and build such a Turing machine **N**, which given a word, produces its number. Having machines **A, D, N** and **U**, we design such Turing machine **X**, that **D** does not produce its description. As we assume that **D** generates descriptions of all Turing machines, we come to a contradiction. This proves impossibility of machine **A** and thus, unsolvability of the stopping problem.

Here we informally describe functioning of **X** utilizing machines **A, D, N** and **U**. Those who are interested in formalizing these considerations can find an appropriate technique in the book (Ebbinghaus et al, 1970).

Machine **X** contains machines **A, D, N** and **U** as subsystems (procedures). When **X** receives an input word $u$, it uses machine **N** to find the number $n$ of $u$. Then **X** uses machine **D** to produce a description of the Turing machine $T_n$ with number $n$. All Turing machines are enumerated in the sequence as their descriptions are produced by **D**. Then **X** uses machine **A** to find if $T_n$ gives a result being applied to the word $u$. If $T_n$ gives no result, then machine **X** gives the result 0. If **A** informs that $T_n$ gives a result, then **X** uses machine **U** to simulate $T_n$ with the input $u$ finding the result $x$ of applying $T_n$ to $u$. When this result is 1, **X** produces 0. In all other cases, **X** produces 1. Thus, **X** is distinct from $T_n$ because **X**($u$) is different from $T_n(u)$. As the word $u$ is taken arbitrarily, **X** cannot coinside with any of the machines, descriptions of which are produced by **D**.

Thus, we have found a problem unsolvable by Turing machines. Now let us show how some inductive Turing machine **M** solves this problem. Given a word $u$ and description D(**T**) of a Turing machine **T**, machine **M** uses machine **U** to simulate **T** with the input $u$. While **U** simulates **T**, machine **M** produces 0 on the output tape. If machine **U** stops, and this means that **T** halts being applied to $u$, machine **M** produces 1 on the output tape. According to the definition, the result of **M** is equal to 1 when **T** halts and the result of **M** is equal to 0 when **T** never halts. In such a way, **M** solves the halting problem.

So, even the simplest inductive Turing machines are more powerful than conventional Turing machines. At the same time, the development of their structure allowed inductive Turing machines to achieve much higher computing power than have the simplest inductive Turing machines described above. This contrasts such a property of a conventional Turing machine that changing the structure, we cannot get greater computing power.

There are different types and kinds of inductive Turing machines: with structured memory, structured rules (control device), and structured head (operating device). To measure their computing power, we use such mathematical construction as the arithmetical hierarchy of sets. It is possible to find its description in (Rogers, 1987). In the arithmetical hierarchy each level is a small part of next level. Conventional Turing machines compute two first levels of this infinite hierarchy. What is computed by limit recursive and limit partial recursive functions and obtained by inductive inference is included in the fourth level of the hierarchy. The same is true for the trial-and-error machines recently introduced by Hintikka and Mutanen (1998). At the same time, it is possible to build a hierarchy of inductive Turing machines that compute the whole arithmetical hierarchy.

Although the Church-Turing Thesis was refuted as an absolute and universal principle, it is reasonable to search in what conditions the Thesis is valid. In the same way, scientists look for conditions of validity for natural laws. Such validation of the Thesis has to go into three directions: test it for actual computers, verify it for theoretical computing schemes, and examine its consistency for axiomatic theories. For example, this Thesis may be proved in some axiomatic contexts and disproved in others. A relevant context for such studies of the Thesis might be provided by some theory of formal computations like the axiomatic theory of algorithms (Burgin, 1985) or theory of computations on abstract structures (Moschovakis, 1974). For example, choosing appropriate axioms, it is possible to prove the Church-Turing Thesis in the theory of trans-recursive operators (Burgin and Borodyanskii, 1991). One of these axioms states that the result of a computation is obtained after a finite sequence of steps and we know when it happens. Without this axiom, we come to the class of all inductive Turing machines with recursive memory. In some sense, these machines are such super-recursive algorithms that are the closest to the recursive algorithms. More exactly, it is possible to say that inductive Turing machines are the most powerful among those super-recursive algorithms, which lie one step from conventional models of algorithm, and are the most realistic among the most powerful super-recursive algorithms.

**4. From virtual perspectives to actual reality**

Here we consider three questions: how modern computers and networks are related to super-recursive algorithms, what new possibilities open super-recursive computations, and how it is possible to realize these computations technologically. To achieve the last but not the least goal, we need, in our case, to develop a new paradigm for computing.

We begin with the question if the super-recursive approach, which is very powerful theoretically, is what will be achieved only in some distant (if any) future or it is some that we have right at hand but only do not understand it.

To our surprise, we find that people do not see correctly how computers are really doing. An analysis of computer functioning demonstrates that while recursive algorithms (such as Turing machines) gave a correct theoretical representation for computers at the beginning of "computer era", super-recursive algorithms are more adequate as mathematical models for modern computers.

At the beginning of computer era, it was necessary to print out some data to get a result. After printing, computer stopped functioning or began to solve another problem. Now people are working with displays. A computer produces its results on the screen of a display. Those results on the screen exist there only if computer functions. This is exactly the case of an inductive Turing machine because the majority of its results is obtained without stopping. A possibility to print some results and to switch off the computer only shows that recursively computable functions constitute a part of functions computed by inductive Turing machines.

It is useful to understand that misunderstanding with computers is not unique and similar "blindness" is not new in society. For example, people thought for thousands of years that the Sun rotated around the Earth and only in the 16[th] century Copernicus proved that the reality was different.

Let us consider some examples of contemporary computer utilization. One of the important applications of computers is simulation used for prediction. However, no single computer run or computer output can be considered to be a definitive forecast of what will happen. It is necessary to have many simulations resulting in the form of stacks of computer outputs in order to make more or less valid prediction. Consequently, in the sequence of these simulations, there is no, as a rule, such a moment when the researcher who carries out these simulations can stop computer and say, "Here is the final result." Even when some conclusions are made basing on the output data of simulation, it is possible that after some time the researcher repeats simulation procedure one or several times more. The goal of such repetitions is, as a rule, to obtain more exact or adequate results, to achieve better understanding, or to test some hypothesis. This situation evidently demonstrates a conventional algorithm can adequately represent that only one run of computer simulation, while the whole process has a very different nature.

Such big networks as INTERNET give another important example of a situation in which conventional algorithms are not adequate. Network functioning is organized by algorithms embodied in a multiplicity of different programs. It is generally assumed that any computer program, is a conventional, i.e., recursive algorithm. However, a recursive algorithm has to stop to give a result, but we cannot say that a network shuts down, then something is wrong and it gives no results. Consequently, recursive algorithms turn out to be inadequate.

These examples and many others vividly demonstrate why a problem of advancement of conventional models of algorithm has been so essential for a long period of time. The solution was given by elaboration of the super-recursive algorithms. It has enabled elaboration of a new paradigm for computation or, more generally, for information processing. The conventional paradigm is based on our image of computer utilization, which consists of the following stages: 1) formalizing a problem; 2) writing a computer program; 3) obtaining a solution to the problem by program execution. In many cases, the necessary computer program exists and we need only the third stage. After this, you either leave your computer to do something else or you begin to solve another problem.

This process is similar to the usage of a car. You go by car to some place, then possibly to another and so on. However, at some moment, you park the car you were driving at some place, stop its functioning, and for a definite time you do not use it. This is the *Car Paradigm* when some object is utilized only periodically for achieving some goal but after this it does not function (at least for some time).

In a very different manner, people use clocks. After buying, they start the clock, and then the clock is functioning until it breaks. People from time to time look at the clock to find what time is it. This is the *Clock Paradigm* when some object is functioning all the time without stopping, while those who utilize it from time to time get some results from it. Recursive algorithms imply that modern computers are utilized according to the *Car Paradigm*, while super-recursive algorithms suggest for computer utilization the new *Clock Paradigm*.

Normal functioning of modern computers presupposes that they work without stopping. However, many of them are switched off from time to time. In any case, these devices eventually end computations. At the same time, development of computer technology gave birth to systems that include as their hardware many computers and other electronic devices. As an example, we can take the contemporary World Wide Web. These systems possess many new properties. For instance, who can imagine now that the World Wide Web will stop its functioning even for a short period of time? Thus, the World Wide Web is a system that works according to the *Clock Paradigm*. Consequently, only super-recursive algorithms such as inductive Turing machines can correctly model such systems.

Although, some networks are functioning in the Clock Paradigm, conscious application of the new approach provides for several important benefits. First, it gives better understanding of computational results obtained during some finite period of time. Second, it shows how to utilize computers in a better way. Third, it makes possible to use more adequate theoretical models for investigation and development of computers. For example, simulation and/or control of many technological processes in industry is better modeled when these processes are treated as potentially infinite.

Embedded computing devices that employ super-recursive schema will be working in the Clock Paradigm if their host system is stationary. At the same time, embedded computing devices that employ super-recursive schema will be working in the Watch Paradigm if their host system moves. Such host system may be a user, car, plane etc. The same is true for ubiquitous computing. According to its main idea, computations are going on continuously but computing device is not fixed at one place as in the *Clock Paradigm* but moves together with its owner. Recently, such approach as nomadic computations has been coined to reflect new facilities hidden in the Internet. Computers will be connected to Internet all the time and will work without stopping. It does not mean that they will function in the same mode all the time. This gives another example of the Clock Paradigm.

Even Internet computing will be on the lines of the Clock Paradigm because Internet will work without stopping and users enter it like people who look at a clock to know time. Only in this case, it will take more efforts to "look" but it will be possible to get much more information. To transform the contemporary Internet into an actual Cyberspace, we have to provide that it functions without stopping like a clock or nature. Can you imagine that the universe stops functioning? That was not by chance that the great Newton compared nature to a clock. Computers can really transit from a part of our World into our environment only when the new paradigm will be used.

In addition to this, many problems that are now considered undecidable will be repeatedly solved by future computing systems. Those will include such practically important problems as debugging, program optimization and many others.

In (Alt,1997) such problems as weather forecast and medical control and diagnosis are considered. According to the description of future methodology for their solution, computers that will solve them have to work in the Clock Paradigm. In addition to this, new approach to computation will enable usage of better numerical methods. Consequently, weather forecast and medical control and diagnosis will achieve much higher level of reliability than they have now.

It is worth mentioning that this novel approach is possible to implement even with the existent hardware, software, and infware. Although, it will be realized only in a partial form. Current situation is reflected in the following metaphor. People have some primitive planes, but do not know how to use them for flying. A new theory explains how to fly on these planes and how to build much more advanced planes. This implies that people will need new skills, because flying is different from driving even the best car.

To realize the new paradigm to a full extent, we need innovative hardware based on different physical principles (Stewart, 1991), original software implementing super-recursive principles of information processing, and even nonpareil organization of infware.

Now we come to the problem of application of the new paradigm. To utilize its higher possibilities and thus to go beyond the Church-Turing Thesis, people and the whole society have to use their creativity. Even more, it will have to be the creativity of special kind that may be called the grounded or intelligent creativity. Humans and computers have to be cooperating systems (Norman, 1997). Consequently, creativity will continue to be a clue to the highest achievements, but super-recursive computations will increase these heights to unimaginable extent. Thus, people's creativity multiplied by computing power of super-recursive devices and algorithms will cause the real revolution in information technology and life of people.

However, some aspects of creativity might be programmed (Burgin and Povyakel, 1988). Then programmed creativity times super-recursive computing power will give birth to Artificial Intelligence that will really be on the same or even on the higher level than human intellect.

## 5. Conclusion

In business and industry, the main criterion is enterprise efficiency, which is called in economics productivity. Consequently, computers are important not because the make computations with more speed than before. What really matters is that they can increase productivity. However, reaching the goal of higher productivity depends on improved procedures. Without proper procedures and necessary skills of the performers, technical devices can decrease productivity. Consequently, methods that develop computer procedures are more important than improvement of hardware or software. In other words, if we compare the super-recursive approach to quantum and DNA computing, we see that the super-recursive approach has more advantages as it implies procedural changes, while quantum and DNA computing provide only technological innovations. Although, these innovations might be very essential, when realized.

We may compare these three approaches to what we have in our everyday life. From this perspective, DNA computing is like a new model of a car. Quantum computations are like planes at the stage when people did not have them. At the same time, super-recursive computations are like rockets, which can take people beyond the "Church-Turing Earth".

Thus, it is possible to say that we have rockets that might take us to the Moon and even to other planets of the Solar system if we know how to navigate them. However, we will need new physical ideas for realization of super-recursive algorithms to a full extent. Using our metaphor, we may say that spaceships that will take us to stars are now only in perspective.

Cyberspace will be integrated into the physical world on the majority of levels and at all places. Ipsyses will provide information about different natural, technical and social systems, control other systems, and help human beings immensely.

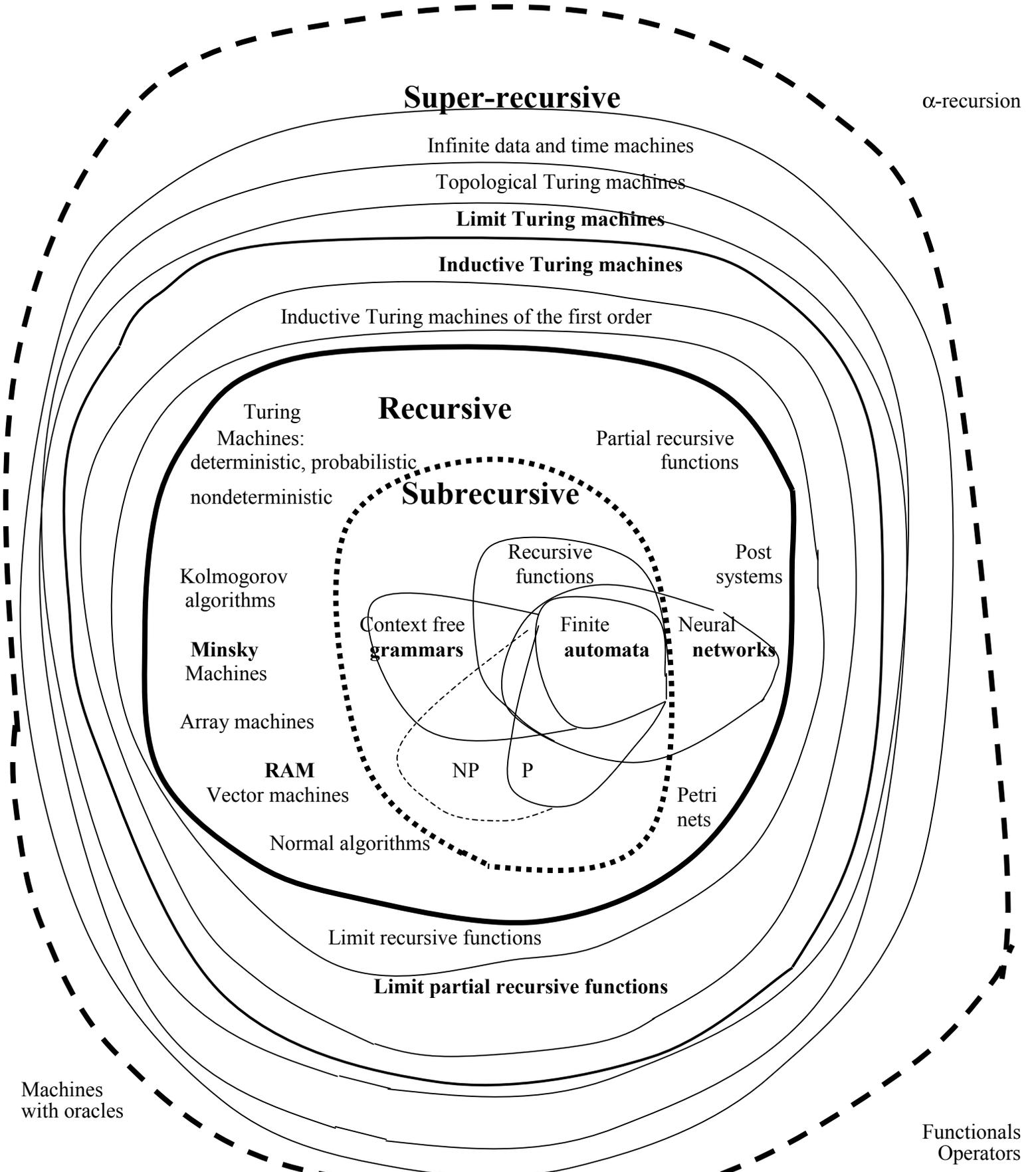

**Fig.2.** Algorithmic Universe

To conclude, it is necessary to remark that there are several models of super-recursive algorithms. The whole picture of the algorithmic universe is given in Fig. 2. The main advantage of inductive Turing machines is that they work with finite objects and obtain the results through a finite period of time. At the same time, other models either work with infinite objects, such as real numbers, or need infinite time to produce the result. Examples of the first type are Turing machines working with real numbers and neural networks working with real numbers (Siegelman, 1999) and some types of topological algorithms (Burgin, 2001). Examples of the second type are infinite time Turing machines (Welch, 2000), persistent Turing machines (Goldin and Wegner, 1998) and some types of topological algorithms (Burgin, 2001). Working with finite words during finite time, these models are either equivalent or weaker than deterministic Turing machines.

Thus, we may say that the good news are that inductive Turing machines give results in a finite time, are more powerful and efficient than ordinary Turing machines and other recursive algorithms. The bad news are that they do not inform when they get their results. However, this property makes them far better models for a majority of real-life systems and processes (cf., for example, Burgin, 1993).

As it is outlined above super-recursive algorithms have important implications for artificial intelligence. Thus, many contrast creativity and formal reasoning. For example, Gregory Chaitin (1999) writes that it is much optimistic to think about mathematical discovery, about creativity instead of formal reasoning... Recursive algorithms, as it is well known, embody formal reasoning. Super-recursive algorithms incorporate genuine creativity. More exactly, they bridge informal creativity and formal reasoning. If we look at the picture of the Algorithmic Universe on the Fig. 2, we see that recursive region of the universe is closed system, which makes depressing incompleteness results, such as Gödel incompleteness theorems, sound like something absolute that is impossible to overcome. In contrast to this, super-recursive region of the Algorithmic Universe is open. It implies development, creativity, and puts no limits on human endeavor.

To conclude, it is worth mentioning that theory of super-recursive computing has an important message for society and not only for technology. It only looks smart to stop and to enjoy after you get something. Theory of super-recursive computing says that it is be better to continue to be active. This is the truth of life.